# IRS-aided UAV for Future Wireless Communications: A Survey and Research Opportunities

ANAS ALKHATIEB [1], KHALED RABIE [2], XINGWANG LI [3], GALYMZHAN NAURYZBAYEV [4], AND RAMEZ ALKHATIB [5]

[1] Department of Computer Engineering, Umm Al-Qura University, Macca, Saudi Arabia
[2] Department of Engineering, Manchester M University, Manchester, UK
[3] Department of Physics and Electronic Information Engineering, Henan Polytechnic University, China
[4] School of Engineering and Digital Sciences, Nazarbayev University, Astana, Kazakhstan
[5] Department of Computer Science, Hama University, Hama, Syria

Corresponding authors: Anas Alkhatieb(s43880520@st.uqu.edu.sa), Khaled Rabie( k.rabie@mmu.ac.uk), Xingwang Li (lixingwangbupt@gmail.com), Galymzhan Nauryzbayev (galymzhan.nauryzbayev@nu.edu.kz) and Ramez Alkhatib ( ramez.alkhatib@hama-univ.edu.sy)

**Abstract—** Both unmanned aerial vehicles (UAVs) and intelligent reflecting surfaces (IRS) are gaining traction as transformative technologies for upcoming wireless networks. The IRS-aided UAV communication, which introduces IRSs into UAV communications, has emerged in an effort to improve the system performance while also overcoming UAV communication constraints and issues. The purpose of this paper is to provide a comprehensive overview of IRS-assisted UAV communications. First, we provide five examples of how IRSs and UAVs can be combined to achieve unrivaled potential in difficult situations. The technological features of the most recent relevant researches on IRS-aided UAV communications from the perspective of the main performance criteria, i.e., energy efficiency, security, spectral efficiency, etc. Additionally, previous research studies on technology adoption as machine learning algorithms. Lastly, some promising research directions and open challenges for IRS-aided UAV communication are presented.

## I. INTRODUCTION

Intelligent reflective surface (IRS), a recently discovered artificial material with electromagnetic properties , for example, refraction, reflection, absorption, and phase, that can be electronically configured in real-time is shown in Fig. 1. These surfaces can be produced at a lower cost, allowing them to be deployed universally and providing the ability to control the wireless multipath environment both outdoors and indoors [1]. Therefore, there's a different way whereby IRS reveals the fields of wireless studies of which the attention focuses on addressing multipaths to designing it. Researchers lately investigated that IRS-aided solutions may greatly improve present vehicular and mobility networks' energy efficiency capacity, and coverage [2], [3], [4], [5], -[6]. In this regard, different types of metasurfaces are as follows:

- Programmable/-reconfigurable metasurface and Large Intelligent Metasurface (LIM) [7], where the property found in naturally occurring surfaces are designed to have some features which are unnaturally occurring surfaces.

- Large Intelligent Surface (LIS) [8], [9], which includes large arrays of low-cost antennas that are generally spaced half of the wavelength apart, It should be mentioned that, in [10], [11], the surface is actively transmitting signals, rather than passively reflecting signals.

- Smart Reflect-Arrays [12], [13], that is based on the surface's reflection function (similar to IRS), have the function of transmission furnished by amplifying and-forward relaying (AF) for instance multiple input, multiple output (MIMO) Technik.

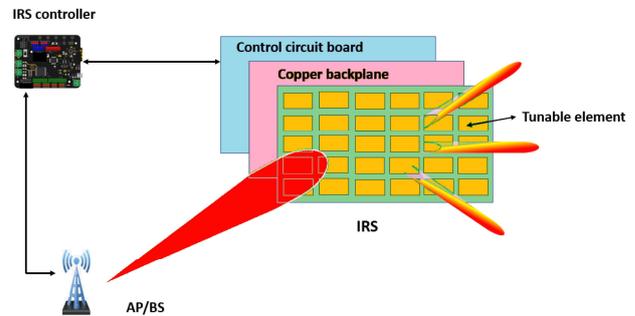

*Fig 1. IRS takes an incoming wave and reflects it as a beam in a particular direction (or towards a spatial point)*

Passive intelligent mirrors and [14] Passive Intelligent Surface (PIS) [15], do not consume any transmit power due to their nature of passive reflection.

- Software-Defined Surface (SDS) [16], [17], encouraged through the definition of software-defined radio regard interaction among the surface and received waves to have a software-controlled metasurface.

Reconfigurable Intelligent Surface (RIS) [18], [19], [20], [8], and [21], where "reconfigurable" relates to the control of reflection angle (by software) irrespective of the angle of incidence. The designs of IRS from several teams are diverse, but they all follow several characteristics as follows:

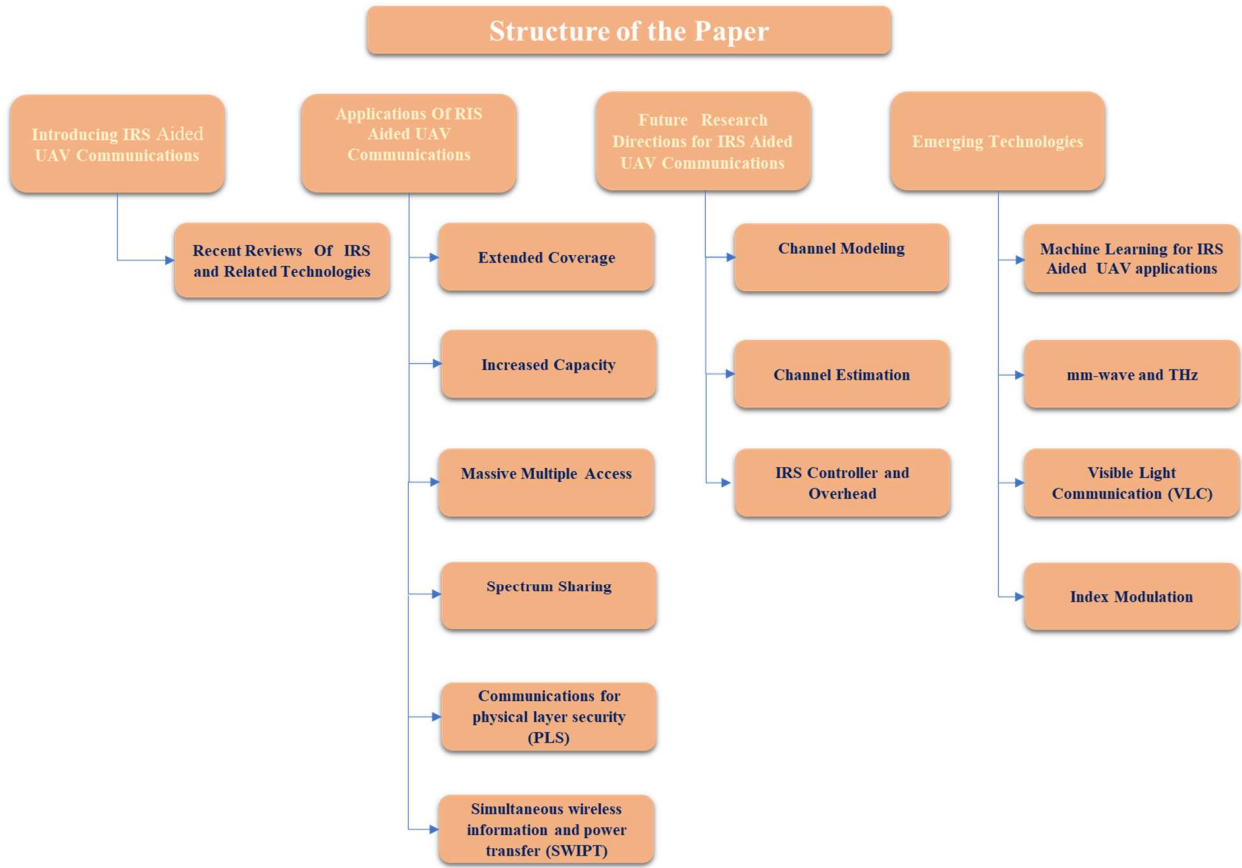

Fig. 2. The overall organization of the paper: Section II defines emerging wireless communications and several use cases for RIS-aided UAV. Section III captures the challenges and open problems, whereas Section IV discusses the role of the IRS and next-generation wireless technology. Section IV derives the conclusions.

- **Nearly Passive**: Instead of expanding the reflecting signals, IRSs reconfigure the channels by regulating reflections. The power needed to program and control phases is quite low. During or after configuration, there is no analog-to-digital or digital-to-analog conversion or power amplification.
- **Reconfigurable**: Reflecting elements are individually regulated to change the phase, amplitude, frequency, and polarization of incident radio waves. The entire surface is viewed as contiguous, with the possibility of changing the shape of the signals at any time and the requirement for real-time reconfigurability.
- **Easy deployment**: IRSs are 2D surfaces with thin layers that are easy to be deployed. They are simple to install on windows, ceilings, and even building exterior walls.

Unmanned Aerial Vehicles (UAVs) have distinct advantages in this regard, for example, greater degrees of freedom in trajectory and location, maintenance costs, less deployment, and the adaptation to determine clear line-of-sight (LOS) [22]. In the 3rd 3GPP (Generation Partnership Project), drones can function as aerial user equipment (AUE) or an aerial relay (AR) that co-exists with terrestrial users, or it forms a portion of wireless infrastructure offering a range of services to the wireless networks. The flowchart in Fig. 2. shows the paper's structure and organization.

## II. APPLICATIONS OF IRS ASSISTED UAV COMMUNICATIONS

Multiple networking and communications rely on the advantageous integration of IRSs and UAVs. In this paper, we disclose the impacts that this framework can have on physical layer security (PLS), massive multiple access (MA), coverage as well as simultaneous wireless information and power transfer (SWIPT).

### A. Extended Coverage

The aerial base station (ABS) and the aerial relay (AR) solutions are responsible for offering coverage that is vital and flexible in upcoming cellular network distribution. However, active aerial communications, on the other hand, incur significant energy overhead. The IRS is the passive alternative. The intelligent omni- surface (IOS), is a type of IRS that consists of antenna units on each side of the IOS and may reflect incident signals from different directives, allowing it to cover affected areas, enhance spectral efficiency and offer enormous 360-degree access. UAV offers this ability by installing the IOS beneath it and flying at appropriate altitudes availing reflective surfaces where they are required. The IOS has the important capability of managing the orientation of the departure signal by modifying phase shift vectors on each side, from the IRS to receivers without any blind spots. Improving phase shift vectors and UAV trajectory for reaching the intended UE position, both static and movement, successfully expands the authentic cell coverage in a direction that is mostly desired as shown in Fig .3. This IOS mechanism can be thought of as a significant feature in extending their downlink communications reach of clusters of UEs and ground base stations to serve the spread of ground UEs.

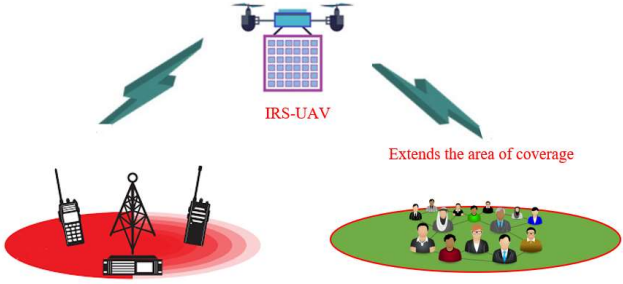

Fig. 3. RIS aided UAV extends the area of coverage of the base station.

### B. Increased Capacity

The rate among UEs and ground BSs on both downlink and uplink can be increased with the positioning of AR. For the UAV's location, direction, transmit energy and various optimizations can be used. The IRS-UAV is an alternative method for increasing throughput and capacity. The IRS typically operates in the full-duplex (FD) method, and when compared to half-duplex (HD) the spectral efficiency is more than the HD method popularly used for ARs. Furthermore, the IRS's passive nature provides for an approach that succeeds in self-interference as well as any antenna noise amplification, resulting in decreased power consumption and less computation than active FD relays [23]. The IRS controller and interference cancellation, according to the channel state information (CSI), can be achieved via managing a few units of the IRS to inverse interfering signal with a result of removing or weakening it. Furthermore, by enhancing the phase shifts of the antenna units, the IRS can work beside the UAV in introducing a rich scattering of LoS links for multiple ground UEs. As a result of UAV-enabled LoS abilities, and the features of the IRS, existing solutions will be exceeded by spectral efficiency to expand capacity. Several UAVs can assist with reaching Scalability, together with fixed IRS based on availability. It is important to note that LoS links are characterized by great channel quality high signal-to-noise ratio (SNR), integrated spatial multiplexing, and multi-user MIMO assisted by the improvement of spectral efficiency through aerial-IRS.

### C. Massive Multiple Access

The number of Internet-of-things (IoT) devices is predicted to exceed 75 billion by 2025, which would cause scarcity regarding bandwidth resources [24]. As a result, providing huge access to IoT devices to link, communicate, and share data has become a difficult task. The growing urbanization complicates this effort since there is a high amount of small-scale IoT devices which have a high probability of being deployed in both indoor and outdoor locations [25], with shadowing and the lack of dependable links posing a barrier in both cases. The challenges of massive access could be adequately achieved by integrating IRS technology with UAV dynamics, ensuring maximum system capacity through optimization of the IRS phase shift vectors.[26]. Indoor wireless channels are steered by IRS-aided communications systems in favor of users who have unique requirements. With IRS-aided UAV communications systems, outdoor virtual reality (VR) applications can be expanded. The three key issues facing outdoor and indoor Virtual Reality users are projected to be interference from neighboring VR devices, energy consumption as huge data transfers, and multi-link communications [27]. To address the challenges, systems involving IRS communication precisely UAVs make up an important component of them. This strategy is assumed where the authors regard phase shift vectors with joint optimization. A strategy such as this is employed in [28] whereby cooperative optimizing the beamforming vectors and the phase shift design of the IRS, trajectory,and UAV height is considered by authors to improve coverage, and capacity as well as provide huge connectivity.

### D. Spectrum Sharing

By diagonalizing the channel matrix, IRS can considerably minimize the interference in environments where devices are sending at the same time in the same frequency band. Therefore, IRS qualifies as a unique feature for enabling spectrum sharing. To reduce the interference to primary users, traditional spectrum sharing is often used as a technique to engage cognitive radios (CR), demanding more reliable and efficient spectrum sensing methods [29]. Spectrum sensing, on the other hand, comes at a cost in terms of energy, and under complex channel conditions, reliability can be compromised.

UAV systems' energy efficiency is critical for long operating times. As a result, with the use of spectrum sharing, IRS-aided UAV systems can increase system capacity in hotspots. The usefulness of IRS systems helping spectrum sharing in indoor contexts has been analytically demonstrated in [30], where aptitude is increased by permitting multiple access in interference and Spectrum sharing between users is reduced by adjusting IRS phase shifters. In [9], authors propose an IRS-aided spectrum sharing technique to boost secondary users' (SUs) capacity while maintaining primary users' (PUs) quality of service (QoS) via phase shift optimization to reach channel diagonalization. A logical extension of this research is spectrum sharing facilitated by IRS-aided UAV, in which the parameters relating to the UAV technology will have a key character in improving the total performance of the wireless networks in real-time, as seen in Fig. 4. After installing IRS on UAVs, it will be necessary to investigate how coordinates, and influence IRS performance of phase shift to increase system capacity.

### E. Physical layer security (PLS )

The use of UAVs to improve the PLS of terrestrial cellular networks has been proposed. The dominating LoS linkages that can be created among ground nodes and an aerial enable this. For example, UAVs can play as an AR among authorized users to increase transmit power while lowering the data rate for the eavesdroppers Furthermore, UAVs can play an important role as friendly jammers via Sending artificial noise (AN) to potential attackers while protecting legitimate users' privacy and data. In typical wireless situations, the aforementioned UAV functions for improving PLS have demonstrated significant potential [31]. Wireless attacks and threats, instead, have been designed to make complex and difficult conditions that factors can affect Wi-Fi performance even when the recommended safeguards are in place. An eavesdropper, for example, can carefully position itself to get a high SNR, possibly more than the destination node. To combat intelligent attackers, IRSs installed on UAVs can be used. According to a previous study, genuine users' secrecy rate increases when the distance between communication peers decreases. The distance between a certain user and the transmission source location can be minimized thanks to UAVs' free movement concept. The IRS phase shifts can then be adjusted so that the original, reflection signals combine beneficially at the authorized user, enhancing the SNR. Some IRS reflecting units, on the other hand, can generate a destructive reflected signal by using different phase shifts to diminish the received SNR at specified areas and reduce the likelihood of eavesdropping. An example of this is shown in Fig. 5.

## III. SIMULTANEOUS WIRELESS INFORMATION AND POWER TRANSFER (SWIPT) TECHNOLOGY

Low-power devices, such as sensors with limited battery life and IoT devices, have recently been recharged using improved wireless energy transfer technology. This can be done by capturing Radio frequency (RF) energy from the environment. The SWIPT idea, in which power and data can be exchanged via electromagnetic (EM) waves for energy harvesting or information decoding (ID), has been extended from the concept of power harvesting or energy harvesting (EH) [32]. time switching (TS) and power splitting (PS) are two prevalent SWIPT techniques (TS). In the power domain, the PS technique separates the received signal into two parts: one for information transfer and the other for EH. In distinct time windows, TS alternates between energy harvesting and data transport. New concept technology for UAV has been established, to aid the SWIPT, particularly when EH devices or ID are spread in difficult situations or crisis circumstances. It is because of their unusual mobility, ease of deployment, and low cost, that UAVs are becoming increasingly popular. As a result, by flying to the nearby location from EH devices or the ground ID, UAVs can optimize the weighted sum rate . Furthermore, employing UAVs to transmit ID and EH signals eliminate the near-far problem that happens when EH devices or ID are dispersed over a vast zone with a huge distance among ground devices. To bridge the gap between terrestrial devices and the AP, UAVs can act as mobile access points (APs). Additionally, IRS-aided UAV SWIPT technology will increase EH and ID for a variety of ground device distributions. In addition to the original EH or ID beams, IRS-aided UAV SWIPT technology will allow the employment of additional beamforming passive beams to maximize WSR. To improve the EH and ID performance, the IRS controller can be utilized to build passive beamforming employing reflecting phase shifters. In Fig 6, an IRS-aided UAV is used to shorten the distances and increase SWIPT performance for EH and ID. EH, and ID is enhanced further by additional options provided by IRS-aided UAV SWIPT. These include the joint optimization of the 3D These options include optimizing the UAV's 3D trajectory and phase shift vectors, the IRS-aided UAV's EH and ID performance and energy consumption, and the number of IRS components of the serviced ID and EH users. Defining the UAV positions for each time interval, as well as the ideal places for charging and information receipt or transfer, are all part of the trajectory in 3d planning and analysis. By reducing the power consumed by the UAV-empowered SWIPT and enhancing transfer power from the flying drones. to the ground nodes that are located away from the electric grid, the best design of the 3D trajectory drive increases the framework's power efficiency. SWIPT use cases enabled by UAVs and supplemented with 3D trajectory optimization would assure EH and ID fairness between the restricted battery resource of ground nodes while minimizing the near-far problem's negative effects. Aligned energy and information beams will be produced by the UAV's beamforming optimization to maximize energy and information transfer capacity.

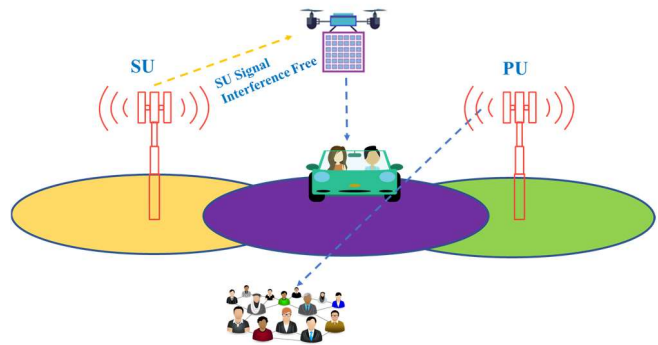

Fig. 4. System model for IRS-UAV underlay spectrum sharing

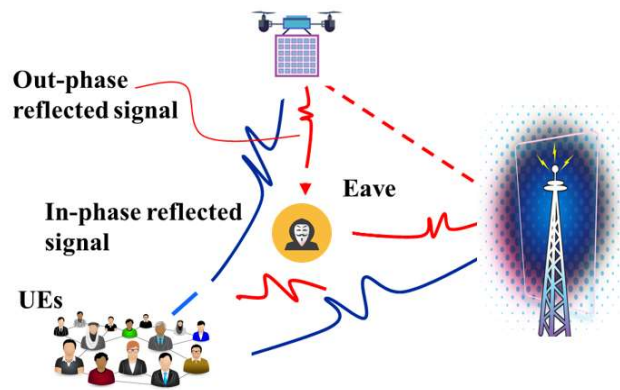

Fig. 5. Illustrations of security issues in IRS-UAV communication systems

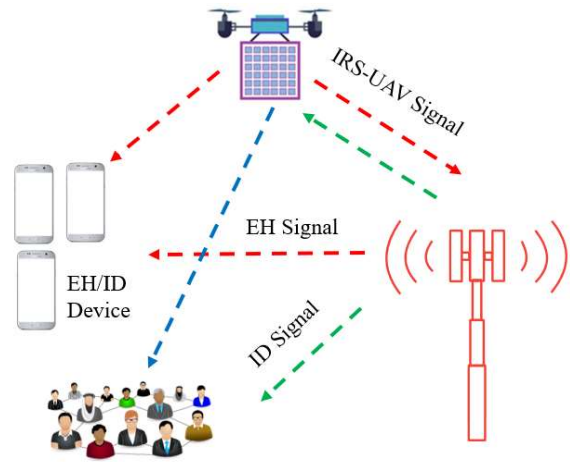

Fig. 6. An IRS-aided SWIPT system comprising one multi-antenna base station (BS).

The coverage radius of the mobile charge air node is enhanced by IRS-backed UAV technology. This enables EH or ID in a variety of scenarios, including dense deployments, large geographic distributions, harsh traffic lights, and disaster situations. In addition, a brief comparison of the IRS-aided UAV networks and the advantages over existing technologies, are summarized in Table 1:

| Metric | Integration of aerial nodes | Aerial IRS-aided networks |
|---|---|---|
| Coverage | • Aerialcommunications nodes can provide additional coverage when and where it is required.<br>• A constellation of UAVs must be formed for broad coverage, which consumes energy and is costly. | • Aerial- IRS enhances coverage by directing the beam toward UEs and allows the UAV to fly about freely while tracking and following target UEs.<br>• A component of the IRS can also be used to charge the UAV's internal battery. |
| Capacity | • To maximize channel capacity, 5G aerial systems employ LoS-MIMO.<br>• Ground users and other aerial nodes may be exposed to severe pilot contamination as a result. | • Aerial-IRS systems tak advantage of the capability of both ground and aerial vehicles and boost them by defining the right number of parts to address issues like pilot contamination. Furthermore, the many degrees of freedom allow for greater spectral efficiency. |
| PLS | • Bytransmitting AN, 5G aerial devices create safe zones for ground users. Jamming and spoofing can be prevented or mitigated using a variety of communication strategies. These prevention strategies may not be able to withstand the limited energy available to aerial systems. | • theseese defenses may not be able to stand up against the limited energy available to aerial systems. Because passive phase shifting is used, significant energy savings are possible. |
| Massive Access | • To service many devices, aerial-aided networks can be dynamically established closer to the end users.<br>• Scalability is harmed by UAV spectrum access and interference. | • An aerial IRSreduces interference by reflecting signals to the intended users.<br>• Multi-UAV-IRS systems can be coordinated with static IRSs to achieve scalability. |
| Spectrum Sharing | • aerialal aided networks give spectrum sharing additional flexibility, for example, by minimizing the interference footprint by smart UAV location. | • withith the use of IRS, aerial IRS systems provide flexible spectrum sharing, allowing several users to share spectrum without interfering with one another. |
| SWIPT | • theThe near-far problem can be addressed by the 3D mobility pattern of UAVs and LoS connectivity. Furthermore, aerial nodes are critical enablers for providing SWIPT, particularly in areas where terrestrial networks have been degraded. The restricted flight time of aerial nodes is a problem, particularly when serving dispersed consumers. | • Aerial IRS-aided systems deliver EH and ID beams alongside aerial or terrestrial network signals so that multiple users can be served simultaneous. The capacity to employ a portion of the IRS elements to continuously charge the aerial nodes though the user-placed SWIPT duties are in progress will lengthen the period of the system. |

### IV. CHALLENGES AND OPEN PROBLEMS

Many superior works are appearing recently, but there are still many challenges that lead to future research directions. In this section, we suggest potential areas of research and future directions.

#### 1. Channel Modeling

Both basic and applied research involves channel modeling, which requires trustworthy data to quantify path loss, shadowing, scattering, and fading effects, among other things. When building accurate models of IRS-aided UAV channels, several things must be considered. The aerial and ground distances, IRS construction material, the sum of pieces, and IRS geometry are all factors to consider. The ability to change the behavior of a communication channel and its impact in various usage scenarios is investigated theoretically and through simulations using theoretical channel models. The IRS-aided UAV channel is made up of two key components: the UAV and the IRS, which together make channel modeling complex and complicated. A UAV will produce a wide range of spatial and temporal alterations as an aerial node with total freedom of movement and aerial shadowing that is varied by motion. As a result of its reflective and passive behavior, as well as the near-field zone propagation that must be considered, the IRS complicates the definition of appropriate channel models [33]. Fig. 7 outlines some critical challenges for channel estimation of IRS-assisted UAVs.

### 2. Channel Estimation

Optimizing the phase shift vector as a purpose of the channel among radio and IRS is a common way of determining the IRS's performance. As a result, determining the channel among the serving radios and the IRS elements is crucial for optimal beamforming and radio channel control. Since no power amplifiers or data converters are required, the IRS elements' passive nature identifies this technology as low-complexity and energy-efficient. The channel's passive character, on the other hand, makes estimating more difficult. To address this, a variant of the passive IRS configuration is proposed. A few low-power active sensors can be used in conjunction with IRS elements to sense and estimate the radio channel. If these active units are integrated with the IRS, they can be used to send and receive pilot signals to obtain the accuracy of available CSI. Instead of channel estimation, machine learning (ML) methods, notably reinforcement learning, can be used as alternative Reinforcement Learning (RL). It is possible to use a Markov Decision Process (MDP) to solve a problem, the mathematical framework of most RL methods, is frequently used to simulate channel states, and RL can interact with the channel to optimize the phases in a computationally efficient manner [34]. In summary, research on the application of ML techniques in IRS-aided UAV communications is still in its initial stage. Nevertheless, the previous works introduced above confirmed that ML is promising to enhance several parameters of complex IRS-aided UAV communications.

### 3. IRS Controller and Overhead

It is crucial to consider how to manage IRS components. This controller sends the phase shift vector to the antenna element. The required phase shift is usually stored in the IRS controller's memory, although this solution needs a properly synchronized and reliable control connection between the IRS and the compute node. For fixed applications and other cases where control connections are readily available, this is possible. However, in the IRS-UAV system, changing channel conditions influence the control link between the compute node and the IRS, causing fading and shadows and interfering with the process of uploading phase shift adjustments in real time. Furthermore, IRS pieces might range in size and number from a few to hundreds or more, resulting in substantial signaling overhead. As a result, new solutions are required to provide low-latency, reliable control links while simultaneously reducing control signaling and processing overhead without sacrificing speed. To assure the stability and availability of control links, one viable approach is to deploy swarms of UAVs that provide distributed computation and communication capabilities.

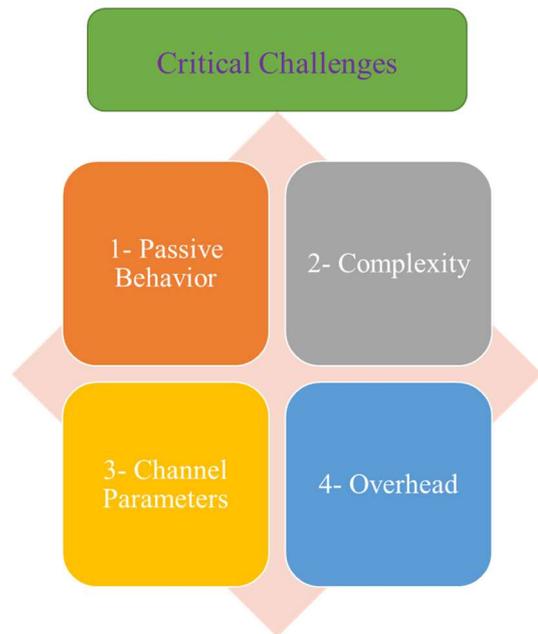

Fig. 7. Crucial challenges in IRS channel estimation.

In Table 2, We have summarized several existing research papers on IRS-aided UAV communications.

| Reference | Channel Model | Design Characteristics | Metric |
|---|---|---|---|
| [35] | Line-of-sight | UAV trajectory and velocity, IRS phase shift | Average total energy consumption |
| [36] | Rician | UAV trajectory, IRS phase shift matrix, IRS scheduling | weighted sum BER minimization |
| [37] | Rician, LOS, Raleigh | UAV trajectory, IRS phase shift, linear precoding | Signal-to-noise-ratio (SNR) |
| [38] | free space path loss, Rayleigh | UAV trajectory, linear precoding, IRS phase shift | Average capacity, average BER, Outage probability |
| [39] | 3GPP | UAV trajectory, IRS phase shift, precoding | Power consumption minimization |
| [40] | Rician | UAV trajectory, IRS phase shift | Average energy consumption |
| [41] | Multipath channel | UAV trajectory, IRS phase shift, precoding, user scheduling, analogue beamforming, | Sum rate |
| [42] | mmWave channel | UAV trajectory, IRS phase shift, precoding | Sum secrecy rate |

## V. RESEARCH OPPORTUNITIES

The upcoming 6G Wi-Fi networks will deliver a continual stream of cutting-edge technology to suit the needs of the future Wi-Fi services and applications. As a result, key characteristics of RIS-UAVs for integration into 6G networks are being investigated [33]. IRS-aided UAVs offer a highly configurable approach to extending the two-dimensional network model to the three-dimensional model, satisfying the requisites of future wireless networks. It can provide reduced energy consumption, secure transmission, high reliability, extended coverage, enhanced QoS, and tuned channel gains. Future networks with the aforementioned benefits will provide a continual stream of breakthrough innovations to meet the needs of cutting-edge applications and services. There are multiple networking and communication use cases where the coexistence of UAVs and IRSs can be advantageous. Here, we disclose the impacts that this framework can have on machine learning (ML), mmWave, THz, Reflection Efficiency, Index Modulation, and Power Consumption. In this section, we discuss the key issues that need to be addressed urgently.

- *ML for IRS-aided UAV applications*

Artificial intelligence (AI) techniques, or most especially ML techniques, can help to enhance network performance, reliability, and QoS Further breakthroughs for the discussed applications are likely when AI and ML are used to empower intelligent IRS-assisted UAV technology. Intelligent drones with IRS capabilities can make autonomous judgments, extract, and forecast Supervised learning is a term used to describe learning knowledge and provide near-optimal improvements [43]. performed through labeled data, Unsupervised learning is when you learn without any labeled data and Reinforcement learning refers to the process of learning from real-time data. To increase IRS channel estimates, spectral efficiency, and the balancing of various tradeoffs, ML methods can be applied. For channel estimation, supervised ML can be used to train on historical data. On historical data, supervised ML can be utilized to train for channel estimation. This tagged data can subsequently be utilized to train a supervisory algorithm that can forecast immediate channel conditions reliably and accurately. Due to the UAV's limited energy and flying time, the massive volume of data gathered during the training phase can be perceived as a stumbling obstacle. As a result, deep neural networks can be utilized efficiently to extract key characteristics from data sets while minimizing computing effort and complexity. It can also lower the computing load per node by applying distributed ML methods. Rather, the learning process is split into many nodes, either air, ground, or both. ML algorithms have the potential to improve and broaden the scope of existing optimization strategies. Most optimization techniques currently emphasize increasing data rate, energy efficiency, and SNR. ML and AI will be crucial in balancing the numerous trade-offs that come with IRS-enabled UAV performance. Optimization of UAV positioning and trajectory planning, for example, for a given direction and structure of IRS units Furthermore, the internal battery of the UAV, as well as the size, weight, and quantity of IRS units that limit the UAV's flight time can be adjusted in tandem to perform specific missions. Another tradeoff is determining how many IRS units to utilize to reflect the incident signal whereas another section of the IRS is used to charge the UAV's internal battery.

- *mm-wave and THz*

Due to their vast accessible bandwidths of tens of GHz, millimeter (mmWave), and sub-terahertz (THz) frequencies are projected to play a key role in 6G wireless networks and beyond [44]. This will necessitate technology and architecture that differs from those of traditional cellular networks. Environmental conditions for mm-wave and THz communications are harsh, resulting in significant attenuation and molecule absorption for high path losses. Furthermore, utilizing more bandwidth raises the blocking rate of the transmitted signal, affecting wireless service reliability and availability. In short-range LoS conditions, THz communications have shown significant promise thus far. As a result, the proposed UAV technology combined with DIS could be a promising approach for overcoming the aforementioned challenges. Over short distances, a UAV can communicate with a THz-band transmitter. The link status is also monitored by the IRS element to ensure that the outbound link has the best signal path possible.

- *Visible Light Communication (VLC)*

Due to its low implementation costs, ultra-high data speeds, and ability to operate in unlicensed spectrum, visible light communication (VLC) has recently appeared as a strong candidate technicality for next-generation wireless.
The range of limited coverage, signal loss with any small motion among the light-emitting diode (LED) transmitter and photodetector receiver, and the requirement for LoS are only a few of the disadvantages. The IRS-aided UAV can help overcome these problems and boost VLC performance. The configuration of an IRS that supports VLC networks for incident beam management differs from that of a standard IRS. A meta-lens or crystal liquid-based IRS could be used in this type of IRS. Using dynamic artificial muscles and the refractive index, the crystal liquid (FLC) and meta-lens-based IRS may shape the situation of incident light signals [45]. The UAV's free movement pattern enables quick and accurate alignment of the LED and the IRS to achieve LoS. Guarantee. As needed, IRS adjusts the phase shift vector to improve the data rate and extend coverage. The IRS element is designed to modify the channel geometry to match the form of the light. The thickness of the IRS or its index of refraction can be adjusted to alleviate this issue. [45].

- *Index Modulation*

The index modulation (IM) approach is another advanced wireless communication technology that is currently being researched. The acquisition of transmission data using accessible transmission unit indexes, such as the transmission antenna or subcarrier index of the Orthogonal Frequency Division Multiplexing waveform, is the basis of the IM method. Cost savings are now guaranteed with reduced peak-to-peak average power ratio of the transmitted signals, error rate, and complexity as IM schemes are integrated into IRS-supported communications. Therefore, IRS-based index modulation can be used in a variety of system components, including transmitters, receivers, and IRS reflector sections. [16] .Therefore, IRS-based IM schemes can be used in a variety of system components, including transmitters, receivers, and IRS reflector sections. Using IRS-aided IM on the transmitter, on the other hand, increases control signaling overhead and assigns an activated antenna index to optimize performance. Unmodulated carriers for spatial shift keying or modulated carriers for spatial modulation can be used to implement IM schemes [16]. The IRS-aided UAV solution, especially for the RIS-SSK scheme, can be situated close to the transmitter to achieve excellent spectrum efficiency.

- *Power Consumption*

The IRS frequently demands an energy supply because there isn't a power amplifier. On the other hand, UAV conservation is essential due to a lack of battery endurance. This shortcoming has caused a bottleneck in the performance, flight time, and battery life of UAVs. It is possible to think of wirelessly recharging UAVs while they are in flight as a potential solution. wireless power transfer (WPT) procedures can be employed to assure mission continuation, and another UAV can be used to transmit the necessary energy. Researchers should therefore create frameworks and procedures that are appropriate, energy-efficient, and able to be optimized to reduce power consumption without sacrificing the effectiveness of IRS-assisted UAV communication.

- *Environmental Factors*

Most experiments assumed a fixed user position and stable UAV flight. However, UAVs frequently experience erratic transmissions and channel estimation errors that should not be overlooked due to inevitabilities caused by vibration and airflow [46]. As a result, the benefits of beamforming design cannot be fully realized. Environmental factors such as wind and rain can cause UAVs to change course and speed, degrading performance and posing safety risks [47]. The investigation of these environmental influences remains difficult.

- *Reflection Efficiency*

Securing the high reflection efficiency of IRSs is essential for practical deployment. The placement and orientation of the IRS are the primary determinants of reflection efficiency; thus, they should be properly planned [37]. On the other hand, as it makes up for long-distance route loss, enhancing reception performance by employing large IRS is essential. Nevertheless, utilizing big IRS will increase the cost and result in several issues, including size and weight [48]. Making a wise trade-off between the IRS's element count and reception performance is therefore important.

- *Security Vulnerabilities*

UAVs are vulnerable to jamming assaults since they can be quickly detected by optical or radar monitoring. Research on secure transmission technologies is therefore necessary to increase the security of the IRS-assisted UAV communications system [49]. Cooperative jamming and traditional beamforming are excellent antijamming strategies that can reduce the quality of eavesdropping channels [50]. Additionally, the CSI estimation must be secured from and robust to jamming attacks as well as reflect partial CSI on the eavesdropping channel in the PLS system's design [51].

- *Phase Shift Controller At IRS*

Most of the investigated instances for a UAV-assisted RIS system are demonstrated to include RIS being mounted on a wall or in a stationary position. The data transmission between IRS and the computing node will have perfect channel conditions in this situation, where a phase shift computation at IRS is necessary, with no effect from channel fading or latency. The link between the compute node and IRS will encounter time-varying channel conditions and may experience the impacts of fading and shadowing in the scenario when a UAV-mounted IRS is required to follow a similar instance in order to achieve the phase shift changes. The link could have latency when there are many IRS parts. Consequently, research into the analysis

## VI. CONCLUSIONS

First, we outlined several scenarios for which IRSs could be included in UAV communication systems. The IRS-assisted UAV communications systems are a workable option for enabling improved services in future wireless systems due to the variable system configuration. Then, we looked into recent developments in IRS-assisted UAV communications in terms of design objectives including spectrum sharing, energy efficiency, and security. The analysis confirmed the advantages and performance increase that came from combining the IRS and UAV as opposed to the benefits that either the IRS or UAV can provide alone. It has been established that optimizing several variables, including UAV altitude and speed, IRS position, IRS phase shift, and beamforming weights, is crucial to maximizing the gain. We also talked about the optimization strategies used in earlier research, with a special emphasis on ML algorithms. The fact that IRS-assisted UAV communications systems frequently entail a wide range of factors highlights how crucial it is to optimize them to increase system performance. Traditional optimization and ML were introduced jointly as techniques to address the issues. Particular attention was paid to the ML algorithm, which was thought to be better suited for resolving complicated issues. Although earlier studies indicated that IRS-assisted UAV communications have a lot of potentials, there are still many technical difficulties and difficult research problems.

Future research issues that we identified include channel estimation, IRS phase shift, reflection efficiency, UAV energy consumption, security vulnerabilities, and environmental concerns. The accuracy of channel estimate must be increased while simultaneously reducing training overhead and power consumption if performance gains with joint designs of UAVs and IRSs are to be realized. Additionally, the channel estimation should be planned to withstand outside influences like jittering and UAV mobility.

REFERENCES

[1] L. Li *et al.*, "Electromagnetic reprogrammable coding-metasurface holograms," *Nat Commun*, vol. 8, no. 1, p. 197, Aug. 2017, doi: 10.1038/s41467-017-00164-9.

[2] S. Arzykulov, G. Nauryzbayev, A. Celik, and A. Eltawil, "RIS-Assisted Full-Duplex Relay Systems," Jul. 2022, doi: 10.1109/jsyst.2022.3189850.

[3] X. Li, Z. Xie, Z. Chu, V. G. Menon, S. Mumtaz, and J. Zhang, "Exploiting Benefits of IRS in Wireless Powered NOMA Networks," *IEEE Transactions on Green Communications and Networking*, vol. 6, no. 1, pp. 175–186, Mar. 2022, doi: 10.1109/TGCN.2022.3144744.

[4] O. Popoola *et al.*, "IRS-Assisted Localization for Airborne Mobile Networks," *Autonomous Airborne Wireless Networks*, pp. 141–156, 2021.

[5] Z. Chu, Z. Zhu, X. Li, F. Zhou, L. Zhen, and N. Al-Dhahir, "Resource Allocation for IRS Assisted Wireless Powered FDMA IoT Networks," *IEEE Internet of Things Journal*, 2021.

[6] J. Wang, "Spectral efficiency maximization for IRS-assisted wireless communication in cognitive radio networks," *Physical Communication*, p. 101528, 2021.

[7] D. Dardari, "Communicating With Large Intelligent Surfaces: Fundamental Limits and Models," *IEEE Journal on Selected Areas in Communications*, vol. 38, no. 11, pp. 2526–2537, Nov. 2020, doi: 10.1109/JSAC.2020.3007036.

[8] L. Zhang, X. Lei, Y. Xiao, and T. Ma, "Large Intelligent Surface-Based Generalized Index Modulation," *IEEE Communications Letters*, 2021.

[9] M. Jung, W. Saad, and G. Kong, "Performance Analysis of Active Large Intelligent Surfaces (LISs): Uplink Spectral Efficiency and Pilot Training," *IEEE Transactions on Communications*, vol. 69, no. 5, pp. 3379–3394, May 2021, doi: 10.1109/TCOMM.2021.3056532.

[10] "Reliability Analysis of Large Intelligent Surfaces (LISs): Rate Distribution and Outage Probability | IEEE Journals & Magazine | IEEE Xplore." https://ieeexplore.ieee.org/abstract/document/8796421 (accessed Dec. 21, 2021).

[11] S. Hu, F. Rusek, and O. Edfors, "Capacity Degradation with Modeling Hardware Impairment in Large Intelligent Surface," in *2018 IEEE Global Communications Conference (GLOBECOM)*, Dec. 2018, pp. 1–6. doi: 10.1109/GLOCOM.2018.8647606.

[12] Q. Chen, Y. Saifullah, G.-M. Yang, and Y.-Q. Jin, "Electronically reconfigurable unit cell for transmit-reflect-arrays in the X-band," *Optics Express*, vol. 29, no. 2, pp. 1470–1480, 2021.

[13] S. Nie, "Ultra-Massive MIMO Communications in the Millimeter Wave and Terahertz Bands for Terrestrial and Space Wireless Systems," May 2021, Accessed: Dec. 16, 2021. [Online]. Available: https://smartech.gatech.edu/handle/1853/64767

[14] "Achievable Rate Maximization by Passive Intelligent Mirrors | IEEE Conference Publication | IEEE Xplore." https://ieeexplore.ieee.org/abstract/document/8461496 (accessed Dec. 21, 2021).

[15] "Channel Estimation and Low-complexity Beamforming Design for Passive Intelligent Surface Assisted MISO Wireless Energy Transfer | IEEE Conference Publication | IEEE Xplore." https://ieeexplore.ieee.org/abstract/document/8683663 (accessed Dec. 21, 2021).

[16] E. Basar, "Reconfigurable intelligent surface-based index modulation: A new beyond MIMO paradigm for 6G," *IEEE Transactions on Communications*, vol. 68, no. 5, pp. 3187–3196, 2020.

[17] C. Liaskos, A. Tsioliaridou, S. Nie, A. Pitsillides, S. Ioannidis, and I. Akyildiz, "An Interpretable Neural Network for Configuring Programmable Wireless Environments," in *2019 IEEE 20th International Workshop on Signal Processing Advances in Wireless Communications (SPAWC)*, Jul. 2019, pp. 1–5. doi: 10.1109/SPAWC.2019.8815428.

[18] A. Alkhatieb, X. Li, R. Alkhatib, K. Rabie, and G. Nauryzbayev, "Intelligent Reflecting Surface - aided UAV Communications:A survey and Research Opportunities," in *2022 13th International Symposium on Communication Systems, Networks and Digital Signal Processing (CSNDSP)*, Jul. 2022, pp. 362–367. doi: 10.1109/CSNDSP54353.2022.9908061.

[19] B. Yang, X. Cao, C. Huang, C. Yuen, L. Qian, and M. Di Renzo, "Intelligent Spectrum Learning for Wireless Networks with Reconfigurable Intelligent Surfaces," *IEEE Transactions on Vehicular Technology*, vol. 70, no. 4, pp. 3920–3925, 2021.

[20] A. Taha, M. Alrabeiah, and A. Alkhateeb, "Enabling large intelligent surfaces with compressive sensing and deep learning," *IEEE Access*, vol. 9, pp. 44304–44321, 2021.

[21] "Wireless Communications Through Reconfigurable Intelligent Surfaces | IEEE Journals & Magazine | IEEE Xplore." https://ieeexplore.ieee.org/abstract/document/8796365 (accessed Dec. 21, 2021).

[22] A. AlKhatieb, E. Felemban, and A. Naseer, "Performance Evaluation of Ad-Hoc Routing Protocols in (FANETs)," in *2020 IEEE Wireless Communications and Networking Conference Workshops (WCNCW)*, Apr. 2020, pp. 1–6. doi: 10.1109/WCNCW48565.2020.9124761.

[23] A. Sabri Abdalla, T. Faizur Rahman, and V. Marojevic, "UAVs with Reconfigurable Intelligent Surfaces: Applications, Challenges, and Opportunities," Dec. 2020. Accessed: Oct. 20, 2022. [Online]. Available: https://ui.adsabs.harvard.edu/abs/2020arXiv201204775S

[24] X. Chen, D. W. K. Ng, W. Yu, E. G. Larsson, N. Al-Dhahir, and R. Schober, "Massive Access for 5G and Beyond," *IEEE Journal on Selected Areas in Communications*, vol. 39, no. 3, pp. 615–637, Mar. 2021, doi: 10.1109/JSAC.2020.3019724.

[25] T. F. Rahman, V. Pilloni, and L. Atzori, "Application Task Allocation in Cognitive IoT: A Reward-Driven Game Theoretical Approach," *IEEE Transactions on Wireless Communications*, vol. 18, no. 12, pp. 5571–5583, Dec. 2019, doi: 10.1109/TWC.2019.2937523.

[26] "RISMA: Reconfigurable Intelligent Surfaces Enabling Beamforming for IoT Massive Access | IEEE Journals & Magazine | IEEE Xplore." https://ieeexplore.ieee.org/abstract/document/9174831 (accessed Dec. 21, 2021).

[27] S. He *et al.*, "A Survey of Millimeter-Wave Communication: Physical-Layer Technology Specifications and Enabling Transmission Technologies," *Proceedings of the IEEE*, vol. 109, no. 10, pp. 1666–1705, Oct. 2021, doi: 10.1109/JPROC.2021.3107494.

[28] "Reconfigurable Intelligent Surface Assisted UAV Communication: Joint Trajectory Design and Passive Beamforming | IEEE Journals & Magazine | IEEE Xplore." https://ieeexplore.ieee.org/abstract/document/8959174 (accessed Dec. 21, 2021).

[29] Z. Tian, Z. Chen, M. Wang, Y. Jia, L. Dai, and S. Jin, "Reconfigurable Intelligent Surface Empowered Optimization for Spectrum Sharing: Scenarios and Methods," *IEEE Vehicular Technology Magazine*, vol. 17, no. 2, pp. 74–82, Jun. 2022, doi: 10.1109/MVT.2022.3157070.

[30] A. Taneja, S. Rani, A. Alhudhaif, D. Koundal, and E. S. Gündüz, "An optimized scheme for energy efficient wireless communication via intelligent reflecting surfaces," *Expert Systems with Applications*, vol. 190, p. 116106, Mar. 2022, doi: 10.1016/j.eswa.2021.116106.

[31] "UAV-Assisted Attack Prevention, Detection, and Recovery of 5G Networks | IEEE Journals & Magazine | IEEE Xplore." https://ieeexplore.ieee.org/abstract/document/9170267 (accessed Dec. 23, 2021).

[32] G. Nauryzbayev, O. Omarov, S. Arzykulov, K. M. Rabie, X. Li, and A. M. Eltawil, "Performance limits of wireless powered cooperative NOMA over generalized fading," *Transactions on Emerging Telecommunications Technologies*, vol. 33, no. 4, p. e4415, 2022, doi: 10.1002/ett.4415.

[33] N. Agrawal, A. Bansal, K. Singh, and C.-P. Li, "Performance Evaluation of RIS-Assisted UAV-Enabled Vehicular Communication System With Multiple Non-Identical Interferers," *IEEE Transactions on Intelligent Transportation Systems*, pp. 1–12, 2021, doi: 10.1109/TITS.2021.3123072.

[34] A. Taha, Y. Zhang, F. B. Mismar, and A. Alkhateeb, "Deep Reinforcement Learning for Intelligent Reflecting Surfaces: Towards Standalone Operation," in *2020 IEEE 21st International Workshop on Signal Processing Advances in Wireless Communications (SPAWC)*, May 2020, pp. 1–5. doi: 10.1109/SPAWC48557.2020.9154301.

[35] D. Ma, M. Ding, and M. Hassan, "Enhancing cellular communications for UAVs via intelligent reflective surface," in *2020 IEEE Wireless Communications and Networking Conference (WCNC)*, 2020, pp. 1–6.

[36] M. Hua, L. Yang, Q. Wu, C. Pan, C. Li, and A. L. Swindlehurst, "UAV-Assisted Intelligent Reflecting Surface Symbiotic Radio System," *IEEE Transactions on Wireless Communications*, vol. 20, no. 9, pp. 5769–5785, Sep. 2021, doi: 10.1109/TWC.2021.3070014.

[37] "Joint Beamforming and Trajectory Optimization for Intelligent Reflecting Surfaces-Assisted UAV Communications | IEEE Journals & Magazine | IEEE Xplore." https://ieeexplore.ieee.org/abstract/document/9078125 (accessed Oct. 08, 2022).

[38] S. Fang, G. Chen, and Y. Li, "Joint optimization for secure intelligent reflecting surface assisted UAV networks," *IEEE wireless communications letters*, vol. 10, no. 2, pp. 276–280, 2020.

[39] X. Liu, Y. Liu, and Y. Chen, "Machine learning empowered trajectory and passive beamforming design in UAV-RIS wireless networks," *IEEE Journal on Selected Areas in Communications*, vol. 39, no. 7, pp. 2042–2055, 2020.

[40] H. Cho and J. Choi, "IRS-Aided Energy Efficient UAV Communication," *arXiv preprint arXiv:2108.02406*, 2021.

[41] L. Jiang and H. Jafarkhani, "Reconfigurable intelligent surface assisted mmwave UAV wireless cellular networks," in *ICC 2021-IEEE International Conference on Communications*, 2021, pp. 1–6.

[42] "Learning-Based Robust and Secure Transmission for Reconfigurable Intelligent Surface Aided Millimeter Wave UAV Communications | IEEE Journals & Magazine | IEEE Xplore." https://ieeexplore.ieee.org/abstract/document/9434412 (accessed Oct. 08, 2022).

[43] M. A. S. Sejan, M. H. Rahman, B.-S. Shin, J.-H. Oh, Y.-H. You, and H.-K. Song, "Machine Learning for Intelligent-Reflecting-Surface-Based Wireless Communication towards 6G: A Review," *Sensors*, vol. 22, no. 14, Art. no. 14, Jan. 2022, doi: 10.3390/s22145405.

[44] M. Tatar Mamaghani and Y. Hong, "Aerial Intelligent Reflecting Surface Enabled Terahertz Covert Communications in Beyond-5G Internet of Things." TechRxiv, Feb. 01, 2022. doi: 10.36227/techrxiv.17066753.v2.

[45] A. R. Ndjiongue, T. M. N. Ngatched, O. A. Dobre, and H. Haas, "Toward the Use of Re-configurable Intelligent Surfaces in VLC Systems: Beam Steering," *IEEE Wireless Communications*, vol. 28, no. 3, pp. 156–162, Jun. 2021, doi: 10.1109/MWC.001.2000365.

[46] H. Jia, J. Zhong, M. N. Janardhanan, and G. Chen, "Ergodic capacity analysis for FSO communications with UAV-equipped IRS in the presence of pointing error," in *2020 IEEE 20th International Conference on Communication Technology (ICCT)*, 2020, pp. 949–954.

[47] S. Malik, P. Saxena, and Y. H. Chung, "Performance analysis of a UAV-based IRS-assisted hybrid RF/FSO link with pointing and phase shift errors," *Journal of Optical Communications and Networking*, vol. 14, no. 4, pp. 303–315, 2022.

[48] X. Pang, M. Sheng, N. Zhao, J. Tang, D. Niyato, and K.-K. Wong, "When UAV meets IRS: Expanding air-ground networks via passive reflection," *IEEE Wireless Communications*, vol. 28, no. 5, pp. 164–170, 2021.

[49] W. U. Khan *et al.*, "Opportunities for Physical Layer Security in UAV Communication Enhanced with Intelligent Reflective Surfaces." arXiv, Sep. 20, 2022. doi: 10.48550/arXiv.2203.16907.

[50] Y. Li, R. Zhang, J. Zhang, S. Gao, and L. Yang, "Cooperative jamming for secure UAV communications with partial eavesdropper information," *IEEE Access*, vol. 7, pp. 94593–94603, 2019.

[51] H. Zhao, J. Hao, and Y. Guo, "Joint Trajectory and Beamforming Design for IRS-assisted Anti-jamming UAV Communication," in *2022 IEEE Wireless Communications and Networking Conference (WCNC)*, 2022, pp. 369–374.

## BIOGRAPHIES

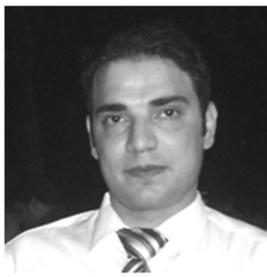

**Mr. Alkhatieb** received a B.Sc. degree (with Hons.) in Information Technology and Computing (ITC) in 2012. subsequently, he received the M.Sc. degree in Computer since and Engineering in 2019 from Umm Al-Qura University in KSA. Additionally, he received the MSc in Strategic Engineering Management from Anglia Ruskin University (ARU) in the UK. He is currently a doctoral student in Electronics and Communication Engineering (ECE). It is a research-based program (RIS-Assisted UAV Toward the Next-Generation Wireless Communication Networks) at Manchester Metropolitan University (MMU) in the UK.

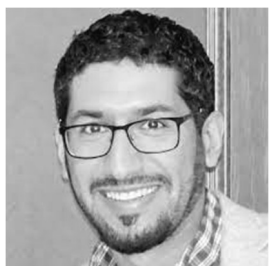

**KHALED M. RABIE** (Senior Member, IEEE) received the M.Sc. and Ph.D. degrees in electrical and electronic engineering from the University of Manchester, in 2011 and 2015, respectively. He is currently a Reader with the Department of Engineering, Manchester Metropolitan University (MMU), UK. He has worked as a part of several largescale industrial projects and has published +200 journal and conference articles (mostly IEEE). His current research interests focus on designing and developing next-generation wireless communication systems. He serves regularly on the technical program committee (TPC) for several major IEEE conferences, such as GLOBECOM, ICC, and VTC. He has received many awards over the past few years in recognition of his research contributions including the Best Paper Awards at the 2021 IEEE CITS and the 2015 IEEE ISPLC, and the IEEE ACCESS Editor of the month award for August 2019. He is currently serving as an Editor of IEEE COMMUNICATIONS LETTERS, an Editor of IEEE Internet of Things Magazine, an Associate Editor of IEEE ACCESS and an Executive Editor of the TRANSACTIONS ON EMERGING TELECOMMUNICATIONS TECHNOLOGIES (Wiley). He guest-edited many special issues in journals including IEEE Wireless Communications Magazine (2021), Electronics (2021), Sensors (2020), IEEE Access (2019). Khaled is also a Fellow of the U.K. Higher Education Academy (FHEA).

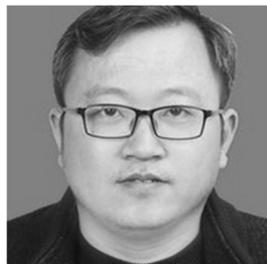

**Dr. Xingwang Li** is an Associate Professor at the School of Physics and Electronic Information Engineering, Henan Polytechnic University, Jiaozuo, China, from Jul. 2015. He was also a Visiting Researcher with Pro. Ping Zhang and Dr. Michail at State Key Laboratory of Networking and Switching Technology, Beijing University of Posts and Telecommunications, Beijing China, and the Institute of Electronics, Communications and Information Technology (ECIT), Queen's University Belfast, Belfast, UK. He received his Master degree and PhD degree in communication and information systems, respectively, from University of Electronic Science and Technology of China, Chengdu, China, and Beijing University of Posts and Telecommunications in Jul. 2010 and Jul. 2015, respectively. Before this, he worked at Comba as an engineer from Jul. 2010 to Aug. 2012.

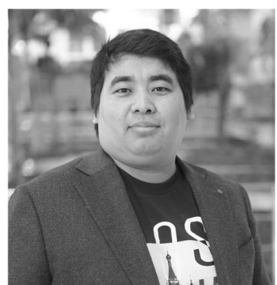

**Dr. Nauryzbayev** received the B.Sc. and M.Sc. degrees (Hons.) in Radio Engineering, Electronics, and Telecommunications from the Almaty University of Power Engineering and Telecommunication, Almaty, Kazakhstan, in 2009 and 2011, respectively, and the Ph.D. degree in wireless communications from the University of Manchester, UK, in 2016. From 2016 to 2018, he held several academic and research positions with Nazarbayev University, Kazakhstan, the L. N. Gumilyov Eurasian National University, Kazakhstan, and Hamad Bin Khalifa University, Qatar. In 2019, he joined Nazarbayev University with the rank of Assistant Professor.

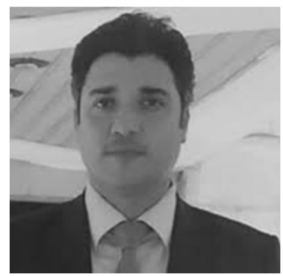

Ramez Alkhatib received the BSc degree in Informatics from University of Aleppo, Syria, the MSc degree in Computer Science from the University of Brunswick, Germany, and the Doctor of Eng. degree in Computer and Information Science from University of Konstanz, Germany. Currently he is an assistant professor of Information Technology at Hama University where he teaches Informatics courses. His current research interests are in Information Technology, especially database management systems.